\pdfoutput=1
\documentclass[11pt,a4paper]{article}
\usepackage[T1]{fontenc}
\usepackage[utf8]{inputenc}
\usepackage{amsmath,amssymb,amsthm}
\usepackage[margin=2.8cm]{geometry}
\usepackage{bm}
\theoremstyle{plain}
\newtheorem{lemma}{Lemma}
\newtheorem{theorem}[lemma]{Theorem}
\newtheorem{proposition}[lemma]{Proposition}
\newtheorem{corollary}[lemma]{Corollary}
\DeclareMathOperator*{\argmin}{arg\,min}
\DeclareMathOperator*{\argmax}{arg\,max}
\newcommand{\R}{\mathbb{R}}
\newcommand{\one}{\bm{1}}
\newcommand{\vzero}{\bm{0}}
\newcommand{\vy}{\bm{y}}
\newcommand{\vb}{\bm{b}}
\newcommand{\vbeta}{\bm{\beta}}
\newcommand{\vz}{\bm{z}}
\newcommand{\hvbeta}{\widehat{\bm{\beta}}}
\newcommand{\tvy}{\tilde{\bm{y}}}
\newcommand{\vu}{\bm{u}}
\newcommand{\hvu}{\widehat{\bm{u}}}
\newcommand{\vv}{\bm{v}}
\newcommand{\lNI}{\lambda_{NI}}
\newcommand{\lS}{\lambda_{S}}
\newcommand{\hvbetaCIS}{\widehat{\bm{\beta}}^{\,CIS}}

\title{A generalized lasso formulation of weighted and regularized\\ (nearly-)isotonic regression on graphs}
\author{Vladimir Pastukhov\thanks{The author started to work on this problem during his postdoctoral position at Chalmers University of Technology, Gothenburg, Sweden, and continued it during his subsequent employment at the University of Vienna, Austria.}}
\date{}

\begin{document}
\maketitle

\begin{abstract}
In this paper we study a regularized version of nearly-isotonic regression
with respect to a general partial order. In this estimator the sum of squares
is weighted, with some of the weights allowed to be zero, and an additional
quadratic penalty is imposed. The only assumption we make is that the
quadratic part of the objective function is strictly convex, and we show that
this assumption is mild. First, we show that for all sufficiently large
values of the penalty parameter the estimator coincides with the regularized
isotonic regression, in which the penalty is replaced by the isotonic
constraint. Further, we prove that the resulting problem can be written as a
generalized lasso problem with a design matrix of full column rank and with
the penalty matrix unchanged. Next, we derive the dual problem, which is a
quadratic program with box constraints, and write it in a sparse form that
can be solved by standard quadratic programming solvers for sparse problems.
\end{abstract}


\section{Introduction}

In the area of constrained inference the basic and simplest problem is
isotonic regression, i.e.\ the $\ell_{2}$-projection of a given vector of
data points $\vy\in\R^{n}$ onto the cone of non-decreasing vectors in
$\R^{n}$,
\begin{equation}\label{eq:iso}
  \hvbeta^{\,I}=\argmin_{\vbeta\in\R^{n}}\|\vy-\vbeta\|_{2}^{2},
  \qquad\text{subject to }\beta_{1}\le\beta_{2}\le\dots\le\beta_{n}.
\end{equation}
A general introduction can be found in \cite{Robertson}. The nearly-isotonic
regression, introduced in \cite{THT} and studied in detail in \cite{Minami},
is the less restrictive version
\begin{equation}\label{eq:ni1d}
  \hvbeta^{\,NI}(\vy,\lNI)=\argmin_{\vbeta\in\R^{n}}
  \tfrac12\|\vy-\vbeta\|_{2}^{2}+\lNI\sum_{i=1}^{n-1}(\beta_{i}-\beta_{i+1})_{+},
\end{equation}
where $(t)_{+}=\max\{t,0\}$, in which the constraint is replaced by a
penalty on its violations. Its combination with the fused lasso and the
extension to a general partial order were studied in \cite{Pastukhov}, where
in particular the dual of the problem was shown to be a quadratic program
with box constraints.

The subject of penalized estimation starts with the $\ell_{1}$-penalisation,
i.e.\ the lasso \cite{Tibshirani1996}, and the $\ell_{2}$-penalisation, i.e.\
ridge regression \cite{HoerlKennard}. The $\ell_{1}$-penalisation of the
successive differences of the coefficients gives the fusion approximator
\cite{Rinaldo}, and its combination with the lasso is the fused lasso
\cite{TibshiraniFL}. These estimators were studied and solved for a general
graph in \cite{Friedman,Hoefling,TT}. In \cite{TT} they appear as special cases
of the generalized lasso, in which the $\ell_{1}$-penalty is imposed on
$D\vbeta$ for a general penalty matrix $D$, and trend filtering on graphs
\cite{WangTF} is another example. The lasso with linear constraints on the
coefficients, called the constrained lasso, and its relation to the
generalized lasso were studied in \cite{Gaines}. There the constraint matrices are assumed to have full row rank, and for isotonic regression this holds only if the graph of the partial order is a tree, which is not the case, for example, for the bimonotone order on a grid. Next, the statistical properties of isotonic
regression in general dimensions were studied in \cite{DengZhang,HanEtAl}, and
algorithms for it can be found, for example, in \cite{BestChakravarti,Stout}.
Several regularized versions of isotonic regression have been proposed.
Generalizations of nearly-isotonic regression in the
one-dimensional case were studied in \cite{YuChenLi}, where a dynamic
programming algorithm is proposed, and in \cite{Gomez}, where the squares of
the positive parts of the differences and a lasso penalty are added to the
penalty in \eqref{eq:ni1d}.

In this paper we study the nearly-isotonic regression with respect to a
general partial order in the following modified form: the sum of squares is
weighted, with some of the weights possibly equal to zero, and a quadratic
penalty is added. This modification is motivated by
\cite{BD}, where the authors study least squares estimation under
bimonotonicity constraints for the data observed on an incomplete grid. There
the weights are the numbers of observations at the grid points, and they are
equal to zero at the empty cells. In this case the problem is not strictly
convex, and the authors add a small quadratic penalty, which they call light
regularization. The quadratic penalty is also the one of the smoothed monotonic
regression of \cite{SysoevBurdakov} in the one-dimensional case. The main
result of the paper is that the modified problem can be written as a
generalized lasso problem in the sense of \cite{TT}, with the same penalty matrix and
with a design matrix of full column rank, and can therefore be solved by the
algorithms developed for the generalized lasso. We also derive the dual
problem.

\subsection{Notation}

We start with basic definitions of partial order and isotonic regression.
Let $I=\{i_{1},\dots,i_{n}\}$ be some index set. A binary relation $\preceq$
on $I$ is called partial order if
\begin{itemize}
\item it is reflexive, i.e.\ $j\preceq j$ for all $j\in I$,
\item it is transitive, i.e.\ $j_{1},j_{2},j_{3}\in I$, $j_{1}\preceq j_{2}$ and
$j_{2}\preceq j_{3}$ imply $j_{1}\preceq j_{3}$,
\item it is antisymmetric, i.e.\ $j_{1},j_{2}\in I$, $j_{1}\preceq j_{2}$ and
$j_{2}\preceq j_{1}$ imply $j_{1}=j_{2}$.
\end{itemize}
Further, a vector $\vbeta\in\R^{n}$ indexed by $I$ is called isotonic with
respect to the partial order $\preceq$ on $I$ if $j_{1}\preceq j_{2}$ implies
$\beta_{j_{1}}\le\beta_{j_{2}}$. We denote the set of all isotonic vectors in
$\R^{n}$ with respect to $\preceq$ by $\mathcal{B}^{is}$. This set is a closed
convex cone in $\R^{n}$, and it is also called the isotonic cone. Next, the
isotonic regression of a vector $\vy\in\R^{n}$ indexed by the partially
ordered index set $I$ is given by
\begin{equation}\label{eq:isogen}
  \hvbeta^{\,I}=\argmin_{\vbeta\in\mathcal{B}^{is}}\sum_{j\in I}(\beta_{j}-y_{j})^{2},
\end{equation}
which for the natural order on $I=\{1,\dots,n\}$ is \eqref{eq:iso}.

For any partial order relation $\preceq$ on $I$ there exists a directed graph
$G=(V,E)$, with $V=I$ and $E$ the minimal set of edges $(j_{1},j_{2})$ with
$j_{1}\preceq j_{2}$, such that $\vbeta$ is isotonic with respect to
$\preceq$ if and only if $\beta_{l_{1}}\le\beta_{l_{2}}$ whenever $E$ contains
a chain of edges from $l_{1}$ to $l_{2}$. We label the vertices by $\{1,\dots,n\}$ and the edges by
$\{1,\dots,m\}$, and let $D\in\R^{m\times n}$ be the oriented incidence matrix
of $G$,
\begin{equation}\label{eq:D}
  D_{e,j}=\begin{cases}
    \phantom{-}1, & \text{if vertex } j \text{ is the source of edge } e,\\
    -1, & \text{if vertex } j \text{ is the target of edge } e,\\
    \phantom{-}0, & \text{otherwise,}
  \end{cases}
\end{equation}
so that $D\vbeta$ is the vector of the differences of $\vbeta$ along the edges
of $G$. Therefore, the isotonic cone is
\[
  \mathcal{B}^{is}=\{\vbeta\in\R^{n}:D\vbeta\le\vzero\},
\]
where the inequality is meant coordinatewise. Further, for $\vz\in\R^{m}$ we
write
\[
  \|\vz\|_{+}=\sum_{e=1}^{m}(z_{e})_{+}.
\]
For the natural order on $I=\{1,\dots,n\}$ the edges join consecutive
indices, and $\|D\vbeta\|_{+}$ is the penalty in \eqref{eq:ni1d}. For the
bimonotone order on a grid, introduced in \cite{BD}, the edges join
horizontally and vertically adjacent grid points.

\subsection{General statement of the problem}

Let $\vy\in\R^{n}$ be a signal indexed by $I$, let $W$ be a diagonal matrix
with nonnegative weights $w_{1},\dots,w_{n}$ on the diagonal, not all equal to
zero, let $S$ be a given matrix with $n$ columns and $\ell$ rows, and let
$\lNI\ge0$. We consider the estimator
\begin{equation}\label{eq:Phi}
  \hvbeta=\argmin_{\vbeta\in\R^{n}}\Phi(\vbeta),\qquad
  \Phi(\vbeta)=(\vy-\vbeta)^{\top}W(\vy-\vbeta)+\|S\vbeta\|_{2}^{2}
  +\lNI\|D\vbeta\|_{+},
\end{equation}
under the assumption that
\begin{equation}\label{eq:A}
  \vbeta\mapsto(\vy-\vbeta)^{\top}W(\vy-\vbeta)+\|S\vbeta\|_{2}^{2}
  \ \text{ is strictly convex.}
\end{equation}
For a weighted graph with nonnegative edge weights $\omega_{1},\dots,\omega_{m}$
the penalty is
\[
  \lNI\sum_{e=(i,j)\in E}\omega_{e}(\beta_{i}-\beta_{j})_{+}=\lNI\|D_{\omega}\vbeta\|_{+},
  \qquad D_{\omega}=\operatorname{diag}(\omega_{1},\dots,\omega_{m})\,D .
\]
Theorem \ref{thm:main} and Propositions \ref{prop:dual} and \ref{prop:lifted}
below hold with $D$ replaced by $D_{\omega}$, since their proofs do not use the
structure of $D$.

The structure of \eqref{eq:Phi} is that of the ``incomplete layout with light
regularization'' in Setting 2b of \cite{BD}. There, the observations of a
bimonotone regression function are reduced to a rectangular grid: $w_{u}$ is
the number of observations at the grid point $u$ and $Z_{u}$ their average,
and the least squares criterion becomes
\[
  \sum_{u}w_{u}(Z_{u}-\theta_{u})^{2}.
\]
When the layout is incomplete, i.e.\ some grid points have no observations
and $w_{u}=0$, this function is not strictly convex, and the least squares fit
is not uniquely determined at the empty cells. For this reason the authors
consider the minimisation of
\[
  \sum_{u}w_{u}(Z_{u}-\theta_{u})^{2}+\lS\,P(\theta)
\]
over the bimonotone cone, where $P$ is a convex quadratic function chosen so
that the sum is strictly convex, and $\lS>0$ is small, its role being only to
make the Hessian well conditioned. The authors discuss two choices of $P$,
the Tychonov regularisation with a reference value $\theta_{0}$ and the sum of
squared differences along the edges of the grid,
\[
  P(\theta)=\sum_{u}(\theta_{u}-\theta_{0})^{2}
  \qquad\text{and}\qquad
  P(\theta)=\sum_{(u,v)}(\theta_{v}-\theta_{u})^{2}.
\]
They prefer the latter, because it interpolates more smoothly into the empty
cells. In the notation of \eqref{eq:Phi} the latter penalty is
$\|S\vbeta\|_{2}^{2}$ with
\begin{equation}\label{eq:BDpen}
  S=\sqrt{\lS}\,D ,
\end{equation}
and their requirement on $P$ is the assumption \eqref{eq:A}. In
Lemma \ref{lem:S} below we verify \eqref{eq:A} for this choice. Problem \eqref{eq:Phi} is thus the
nearly-isotonic relaxation of the problem of \cite{BD}, in which the isotonic
constraint is replaced by the penalty, and the bimonotone order is replaced by
an arbitrary partial order. On the chain graph, with unit weights and with
the constraint kept, the same quadratic part gives the smoothed monotonic
regression of \cite{SysoevBurdakov} with equal penalty parameters.

\subsection{Organisation of the paper}

The rest of the paper is organised as follows. In Section \ref{sec:exact} we
relate the estimator to the constrained estimator in which the penalty
$\lNI\|D\vbeta\|_{+}$ is replaced by the constraint $D\vbeta\le\vzero$. In
Section \ref{sec:red} we show that \eqref{eq:Phi} is a generalized lasso
problem. Section \ref{sec:dual} contains the dual problem. In
Section \ref{sec:S} we verify assumption \eqref{eq:A} for the penalty
\eqref{eq:BDpen}. Section \ref{sec:comp} is devoted to the computational
aspects. The
paper closes with a discussion in Section \ref{sec:disc}.

\section{Relation to regularized isotonic regression}\label{sec:exact}

In this section we compare the estimator \eqref{eq:Phi} with the estimator
obtained when the penalty $\lNI\|D\vbeta\|_{+}$ in \eqref{eq:Phi} is removed
and the isotonic constraint $D\vbeta\le\vzero$ is imposed instead,
\begin{equation}\label{eq:C}
  \hvbetaCIS=\argmin_{\vbeta\in\R^{n}}\ (\vy-\vbeta)^{\top}W(\vy-\vbeta)+\|S\vbeta\|_{2}^{2}
  \qquad\text{subject to}\quad D\vbeta\le\vzero .
\end{equation}
For the grid and $S=\sqrt{\lS}\,D$ this is the estimator of \cite{BD}, and for
the chain it is the smoothed monotonic regression of \cite{SysoevBurdakov}.
For nearly-isotonic regression, i.e.\ for $W=\tfrac12I_{n}$ and $S=0$, the
property that the estimator coincides with isotonic regression for all
sufficiently large $\lNI$ is mentioned in \cite{Minami} and \cite{THT}. The
same holds for the estimator \eqref{eq:Phi}. Under \eqref{eq:A} the objective
functions in \eqref{eq:Phi} and \eqref{eq:C} are strictly convex and coercive,
so both problems have unique solutions. Further, the objective function in
\eqref{eq:Phi} is the exact penalty function of the problem \eqref{eq:C} with
the penalty parameter $\lNI$, and the constraints in \eqref{eq:C} are linear.
Therefore, by the classical exact penalty result, see \cite{Mangasarian} and
the references therein, there exists $\lambda_{0}\ge0$ such that
$\hvbeta=\hvbetaCIS$ for all $\lNI\ge\lambda_{0}$, in particular, for
nearly-isotonic regression with respect to a general partial order.

Thus, the estimator \eqref{eq:Phi} is the regularized isotonic regression
\eqref{eq:C} for all sufficiently large values of $\lNI$, and the results of
the following sections apply to the estimator \eqref{eq:C} as well.

\section{Solution via the generalized lasso}\label{sec:red}

First, we rewrite the objective function in \eqref{eq:Phi}.

\begin{lemma}\label{lem:rewrite}
For a fixed data vector $\vy\in\R^{n}$ indexed by $I$ the objective function
in \eqref{eq:Phi} can be written as
\begin{equation}\label{eq:rewrite}
  \Phi(\vbeta)=\vbeta^{\top}Q\vbeta-2\vb^{\top}\vbeta
  +\tfrac{\lNI}{2}\|D\vbeta\|_{1}+\vy^{\top}W\vy ,
\end{equation}
where
\begin{equation}\label{eq:Qb}
  Q=W+S^{\top}S\in\R^{n\times n},\qquad
  \vb=W\vy-\tfrac{\lNI}{4}\,D^{\top}\one\in\R^{n},
\end{equation}
and $\one\in\R^{m}$ is the vector whose all elements are equal to one.
Further, the assumption \eqref{eq:A} holds if and only if $Q$ is positive
definite.
\end{lemma}

\begin{proof}
The function in \eqref{eq:A} is a quadratic function with constant Hessian
$2Q$, and a quadratic function is strictly convex if and only if its Hessian
is positive definite. This gives the last statement of the lemma.

For \eqref{eq:rewrite}, since $W$ is symmetric, $(\vy-\vbeta)^{\top}W(\vy-\vbeta)
=\vbeta^{\top}W\vbeta-2(W\vy)^{\top}\vbeta+\vy^{\top}W\vy$, and
$\|S\vbeta\|_{2}^{2}=\vbeta^{\top}S^{\top}S\vbeta$. Together these give
$\vbeta^{\top}Q\vbeta-2(W\vy)^{\top}\vbeta+\vy^{\top}W\vy$. For $t\in\R$ one
has $(t)_{+}=\tfrac12|t|+\tfrac12t$. Therefore,
\[
  \lNI\|D\vbeta\|_{+}
  =\tfrac{\lNI}{2}\|D\vbeta\|_{1}+\tfrac{\lNI}{2}\one^{\top}D\vbeta
  =\tfrac{\lNI}{2}\|D\vbeta\|_{1}
  +2\bigl(\tfrac{\lNI}{4}D^{\top}\one\bigr)^{\top}\vbeta,
\]
which finishes the proof.
\end{proof}

Next, we provide the solution to the problem \eqref{eq:Phi}.

\begin{theorem}\label{thm:main}
Assume \eqref{eq:A} holds, let $Q$ and $\vb$ be as in \eqref{eq:Qb}, let
$W^{1/2}=\operatorname{diag}(\sqrt{w_{1}},\dots,\sqrt{w_{n}})$, and let
\begin{equation}\label{eq:Xy}
  X=\sqrt2\begin{pmatrix}W^{1/2}\\ S\end{pmatrix}\in\R^{(n+\ell)\times n},
  \qquad \tvy=XQ^{-1}\vb\in\R^{n+\ell}.
\end{equation}
Then the solution to the problem \eqref{eq:Phi} is given by
\begin{equation}\label{eq:glasso}
  \hvbeta=\argmin_{\vbeta\in\R^{n}}\Bigl\{
  \tfrac12\|\tvy-X\vbeta\|_{2}^{2}+\tfrac{\lNI}{2}\|D\vbeta\|_{1}\Bigr\}.
\end{equation}
That is, the problem \eqref{eq:Phi} is the generalized lasso problem with the
data vector $\tvy$, the design matrix $X$, the penalty matrix $D$ and the
penalty parameter $\tfrac{\lNI}{2}$. Moreover, $X$ has full column rank and
the solution \eqref{eq:glasso} is unique.
\end{theorem}

\begin{proof}
First, note that
\[
  X^{\top}X=2W^{1/2}W^{1/2}+2S^{\top}S=2Q
  \qquad\text{and}\qquad
  X^{\top}\tvy=X^{\top}XQ^{-1}\vb=2\vb .
\]
Next, from Lemma \ref{lem:rewrite} it follows that
\begin{align*}
  \Phi(\vbeta)
  &=\vbeta^{\top}Q\vbeta-2\vb^{\top}\vbeta+\tfrac{\lNI}{2}\|D\vbeta\|_{1}+\vy^{\top}W\vy\\
  &=\tfrac12\vbeta^{\top}X^{\top}X\vbeta-\tvy^{\top}X\vbeta+\tfrac{\lNI}{2}\|D\vbeta\|_{1}+\vy^{\top}W\vy\\
  &=\tfrac12\|\tvy-X\vbeta\|_{2}^{2}+\tfrac{\lNI}{2}\|D\vbeta\|_{1}
  +\vy^{\top}W\vy-\tfrac12\|\tvy\|_{2}^{2}.
\end{align*}
Therefore, the problems \eqref{eq:Phi} and \eqref{eq:glasso} have the same
solutions, since the last two terms do not depend on $\vbeta$. Lastly, by
Lemma \ref{lem:rewrite} the matrix $Q$ is positive definite, so is
$X^{\top}X=2Q$, which means that $X$ has full column rank and the objective
function in \eqref{eq:glasso} is strictly convex and coercive. Therefore, the
solution to \eqref{eq:glasso} exists and is unique.
\end{proof}

The pair $X$, $\tvy$ in \eqref{eq:Xy} is not the only possible choice. As the
proof shows, the statement of Theorem \ref{thm:main} holds for any
$X\in\R^{p\times n}$ and $\tvy\in\R^{p}$ such that
\begin{equation}\label{eq:cond}
  X^{\top}X=2Q\qquad\text{and}\qquad X^{\top}\tvy=2\vb .
\end{equation}
For example, \eqref{eq:cond} holds for the square matrix $X$ and the vector
$\tvy$ given by
\[
  X=\sqrt2\,Q^{1/2},\qquad \tvy=\sqrt2\,Q^{-1/2}\vb .
\]
Finally, for nearly-isotonic regression, i.e.\ for $W=\tfrac12I_{n}$ and
$S=0$, \eqref{eq:cond} holds for
\[
  X=I_{n},\qquad \tvy=\vy-\tfrac{\lNI}{2}D^{\top}\one ,
\]
and \eqref{eq:glasso} becomes the relation
\[
  \hvbeta^{\,NI}(\vy,\lNI)=\hvbeta^{\,F}\bigl(\vy-\tfrac{\lNI}{2}D^{\top}\one,\tfrac12\lNI\bigr)
\]
between nearly-isotonic regression and the fusion approximator, proved in
\cite{Pastukhov}.

\section{The dual problem}\label{sec:dual}

For a fixed $\lNI$ we can also solve the problem \eqref{eq:Phi} numerically through its
dual, which is a quadratic program with box constraints.

\begin{proposition}\label{prop:dual}
Assume \eqref{eq:A} holds. The solution to \eqref{eq:Phi} is given by
\begin{equation}\label{eq:recover}
  \hvbeta=\tvy_{d}-\tilde D^{\top}\hvu ,
\end{equation}
where
\begin{equation}\label{eq:tilde}
  \tvy_{d}=Q^{-1}W\vy\in\R^{n},\qquad
  \tilde D=\tfrac12\,DQ^{-1}\in\R^{m\times n},
\end{equation}
and
\begin{equation}\label{eq:dualprob}
  \hvu\in\argmin_{\vu\in\R^{m}}\ \|\tvy_{d}-\tilde D^{\top}\vu\|_{Q}^{2}
  \qquad\text{s.t.}\quad \vzero\le\vu\le\lNI\one ,
\end{equation}
with $\|\vv\|_{Q}^{2}=\vv^{\top}Q\vv$ for $\vv\in\R^{n}$.
\end{proposition}

\begin{proof}
First, following \cite{TT}, we write the problem \eqref{eq:Phi} in the
following way
\begin{align*}
  &\underset{\vbeta,\,\vz}{\text{minimize}}\quad
  (\vy-\vbeta)^{\top}W(\vy-\vbeta)+\|S\vbeta\|_{2}^{2}+\lNI\|\vz\|_{+}\\
  &\text{subject to}\quad D\vbeta=\vz\in\R^{m}.
\end{align*}
Further, the Lagrangian is given by
\[
  L(\vbeta,\vz,\vu)=(\vy-\vbeta)^{\top}W(\vy-\vbeta)+\|S\vbeta\|_{2}^{2}
  +\lNI\|\vz\|_{+}+\vu^{\top}(D\vbeta-\vz),
\]
where $\vu\in\R^{m}$ is a dual variable. Note that
\[
  \inf_{\vz}\bigl(\lNI\|\vz\|_{+}-\vu^{\top}\vz\bigr)
  =\begin{cases}0, & \text{if } \vzero\le\vu\le\lNI\one,\\ -\infty, & \text{otherwise,}\end{cases}
\]
and, since $Q$ is positive definite,
\begin{align*}
  \min_{\vbeta}\bigl((\vy-\vbeta)^{\top}W&(\vy-\vbeta)+\|S\vbeta\|_{2}^{2}+\vu^{\top}D\vbeta\bigr)\\
  &=-\tfrac14(2W\vy-D^{\top}\vu)^{\top}Q^{-1}(2W\vy-D^{\top}\vu)+\vy^{\top}W\vy\\
  &=-\|\tvy_{d}-\tilde D^{\top}\vu\|_{Q}^{2}+\vy^{\top}W\vy .
\end{align*}
Next, the dual function is given by
\[
  g(\vu)=\inf_{\vbeta,\vz}L(\vbeta,\vz,\vu)
  =\begin{cases}
   -\|\tvy_{d}-\tilde D^{\top}\vu\|_{Q}^{2}+\vy^{\top}W\vy, & \text{if } \vzero\le\vu\le\lNI\one,\\
   -\infty, & \text{otherwise,}
  \end{cases}
\]
and, therefore, the dual problem is
\[
  \hvu\in\argmax_{\vu}\,g(\vu)\qquad\text{subject to}\quad\vzero\le\vu\le\lNI\one,
\]
which is equivalent to \eqref{eq:dualprob}. Lastly, since the objective function is finite and convex on
$\R^{n}\times\R^{m}$ and the constraint $D\vbeta=\vz$ is affine and
feasible, strong duality holds, and taking first
derivative of the Lagrangian $L(\vbeta,\vz,\hvu)$ with respect to $\vbeta$ we
get the following relation between $\hvbeta$ and $\hvu$
\[
  \hvbeta=\tvy_{d}-\tilde D^{\top}\hvu .\qedhere
\]
\end{proof}

If $D$ does not have full row rank, the minimiser in \eqref{eq:dualprob} need
not be unique. However, the objective function in \eqref{eq:dualprob} is a
strictly convex function of $D^{\top}\vu$, therefore, $D^{\top}\hvu$ is
unique, and \eqref{eq:recover} does not depend on the choice of $\hvu$. Note
also that $\tvy_{d}$ is the solution to \eqref{eq:Phi} for $\lNI=0$, so
\eqref{eq:recover} represents the estimator as the quadratic smoother minus a
correction determined by the dual problem. For nearly-isotonic regression we
have $\tvy_{d}=\vy$ and $\tilde D=D$, and Proposition \ref{prop:dual} becomes
Theorem 1 of \cite{Pastukhov} without the fusion penalty.

\section{The strict convexity assumption}\label{sec:S}

In this section we verify the assumption \eqref{eq:A} for the penalty
\eqref{eq:BDpen}, i.e.\ for $S=\sqrt{\lS}\,D$.

\begin{lemma}\label{lem:S}
Let the undirected graph underlying $G$ be connected, and let
$S=\sqrt{\lS}\,D$ with $\lS>0$. Then the assumption \eqref{eq:A} holds if and
only if $w_{i}>0$ for at least one $i\in\{1,\dots,n\}$.
\end{lemma}

\begin{proof}
With $S=\sqrt{\lS}\,D$ we have $Q=W+\lS D^{\top}D$, and by
Lemma \ref{lem:rewrite} the assumption \eqref{eq:A} holds if and only if $Q$
is positive definite. Since $W$ is diagonal and
$\vv^{\top}D^{\top}D\vv=\|D\vv\|_{2}^{2}$, for every $\vv\in\R^{n}$ we have
\begin{equation}\label{eq:vQv}
  \vv^{\top}Q\vv=\sum_{i=1}^{n}w_{i}v_{i}^{2}+\lS\|D\vv\|_{2}^{2},
\end{equation}
and both summands are nonnegative, because $w_{i}\ge0$ and $\lS>0$.

First, let $w_{i}>0$ for some $i$. By \eqref{eq:D}, $D\vv=\vzero$ if and
only if $v_{i}=v_{j}$ for every edge $(i,j)\in E$, and since the undirected
graph underlying $G$ is connected, this holds if and only if $\vv=t\one_{n}$
for some $t\in\R$, where $\one_{n}\in\R^{n}$ is the vector of ones. Now let $\vv\neq\vzero$. If $D\vv\neq\vzero$, the second
summand in \eqref{eq:vQv} is strictly positive. If $D\vv=\vzero$, then
$\vv=t\one_{n}$ with $t\neq0$, and the first summand in \eqref{eq:vQv} equals
$t^{2}\sum_{i=1}^{n}w_{i}>0$. Hence, $\vv^{\top}Q\vv>0$ for all
$\vv\neq\vzero$, i.e.\ $Q$ is positive definite.

Second, let $w_{i}=0$ for all $i$. Then for $\vv=\one_{n}$ we have
$D\one_{n}=\vzero$, because every row of $D$ contains exactly one $1$ and one
$-1$, and therefore $\vv^{\top}Q\vv=0$ by \eqref{eq:vQv}. Hence $Q$ is not
positive definite.
\end{proof}

Therefore, for a connected graph the standing assumption that not all
weights are equal to zero is necessary and sufficient. In particular, for the
bimonotone grid, which is connected, the assumption \eqref{eq:A} holds for
every incomplete layout with at least one observation and for every
$\lS>0$. Note that without the quadratic penalty, i.e.\ for $S=0$, the
Hessian of the quadratic part of \eqref{eq:Phi} is $2W$, which is singular as
soon as $w_{i}=0$ for some $i$.

If the graph is not connected, then for $S=\sqrt{\lS}\,D$ all terms in
\eqref{eq:Phi} are sums over the connected components, because no edge joins
two different components. Therefore, the problem \eqref{eq:Phi} splits into
independent problems, one for each component, and Lemma \ref{lem:S} can be
applied to each of them.

\begin{corollary}\label{cor:components}
Let $S=\sqrt{\lS}\,D$ with $\lS>0$. Then the assumption \eqref{eq:A} holds if
and only if every connected component of the undirected graph underlying $G$
contains a vertex $i$ with $w_{i}>0$.
\end{corollary}

\begin{proof}
Let $C_{1},\dots,C_{K}$ be the connected components of the undirected graph
underlying $G$. Since no edge joins two different components, after a
permutation of the vertices the matrix $Q=W+\lS D^{\top}D$ is block diagonal
with the blocks $Q_{k}=W_{k}+\lS D_{k}^{\top}D_{k}$, $k=1,\dots,K$, where
$W_{k}$ is the diagonal matrix of the weights $w_{i}$, $i\in C_{k}$, and
$D_{k}$ is the incidence matrix of the subgraph of $G$ with the vertex set
$C_{k}$. Therefore, $Q$ is positive definite if and only if every $Q_{k}$ is
positive definite, and the statement follows from Lemma \ref{lem:S} applied to
each component.
\end{proof}

Finally, the Tychonov penalty of \cite{BD} shrinks the estimator towards the
reference value $\theta_{0}$, and for $\theta_{0}\neq0$ it is not of the form
$\|S\vbeta\|_{2}^{2}$. Nevertheless, the corresponding problem
\[
  \min_{\vbeta\in\R^{n}}\ (\vy-\vbeta)^{\top}W(\vy-\vbeta)
  +\lS\sum_{i=1}^{n}(\beta_{i}-\theta_{0})^{2}+\lNI\|D\vbeta\|_{+}
\]
reduces to \eqref{eq:Phi}. Indeed, let $\vbeta'=\vbeta-\theta_{0}\one_{n}$ and
$\vy'=\vy-\theta_{0}\one_{n}$. Then
\[
  \vy-\vbeta=\vy'-\vbeta',\qquad
  \sum_{i=1}^{n}(\beta_{i}-\theta_{0})^{2}=\|\vbeta'\|_{2}^{2},\qquad
  D\vbeta=D\vbeta',
\]
where the last equality holds because $D\one_{n}=\vzero$. Therefore, the
solution is
\[
  \hvbeta=\hvbeta'+\theta_{0}\one_{n},
\]
where $\hvbeta'$ is the solution to \eqref{eq:Phi} with the data vector $\vy'$
and $S=\sqrt{\lS}\,I_{n}$. For this choice of $S$ the matrix $Q$ is positive
definite for any weights, so the assumption \eqref{eq:A} always holds.

\section{Computational aspects}\label{sec:comp}

In this section we discuss the computation of the estimator \eqref{eq:Phi}.
We do not provide a simulation study, because the algorithms below are well
established, and their performance has been studied extensively, see
\cite{Pastukhov} for the nearly-isotonic case, and also \cite{OSQP},
\cite{TT}.

First, by Theorem \ref{thm:main} the estimator is the solution to the
generalized lasso problem \eqref{eq:glasso}. Therefore, we can use the
algorithm of \cite{TT} for the problem \eqref{eq:Phi}, for an arbitrary
partial order, and the condition of full column rank of the design matrix,
required there, is satisfied automatically. Note, that the data vector
$\tvy$ depends on $\lNI$, therefore, the path of \cite{TT} in the lasso
parameter with fixed $\tvy$ is not the path of \eqref{eq:Phi} in $\lNI$.
Further, among all pairs $X$, $\tvy$ which satisfy \eqref{eq:cond}, the choice
\eqref{eq:Xy} is preferable in computations, because the design matrix in
\eqref{eq:Xy} is sparse whenever $S$ is sparse, while $Q^{1/2}$ is dense in
general.

Second, the estimator can be computed through the dual problem
\eqref{eq:dualprob}. However, the matrix $Q^{-1}$ in \eqref{eq:tilde} is dense
in general, even if $Q$ is sparse. To avoid computing $Q^{-1}$, we keep
$\vbeta$ as a variable in the dual problem.

\begin{proposition}\label{prop:lifted}
Assume \eqref{eq:A} holds, and let $(\vbeta^{*},\vu^{*})$ be a solution of
\begin{equation}\label{eq:lifted}
  \min_{\vbeta\in\R^{n},\,\vu\in\R^{m}}\ \vbeta^{\top}Q\vbeta
  \qquad\text{s.t.}\quad Q\vbeta+\tfrac12D^{\top}\vu=W\vy,\quad
  \vzero\le\vu\le\lNI\one .
\end{equation}
Then $\vbeta^{*}=\hvbeta$.
\end{proposition}

\begin{proof}
The constraint $Q\vbeta+\tfrac12D^{\top}\vu=W\vy$ gives
$\vbeta=\tvy_{d}-\tilde D^{\top}\vu$, and then
$\vbeta^{\top}Q\vbeta=\|\tvy_{d}-\tilde D^{\top}\vu\|_{Q}^{2}$. Therefore,
$\vu^{*}$ is a solution to \eqref{eq:dualprob} and
$\vbeta^{*}=\tvy_{d}-\tilde D^{\top}\vu^{*}$, which is equal to $\hvbeta$ by
Proposition \ref{prop:dual}.
\end{proof}

The problem \eqref{eq:lifted} contains only the matrices $Q$ and $D$. Every
row of the incidence matrix $D$ has two nonzero elements, and $Q$ is sparse
whenever $S^{\top}S$ is sparse. In particular, for the penalty \eqref{eq:BDpen}
the matrix $Q=W+\lS D^{\top}D$ has at most $n+2m$ nonzero elements, so it is
sparse when the number of edges $m$ is of the order of $n$, as for the chain
and for the grid. In this case \eqref{eq:lifted} can be solved by quadratic
programming solvers for sparse problems, for example by OSQP \cite{OSQP}, as
it was done for the nearly-isotonic case in \cite{Pastukhov}.

Lastly, if the constrained estimator \eqref{eq:C} is of interest, then, by
Section \ref{sec:exact}, it can be obtained from \eqref{eq:Phi} with a
sufficiently large value of $\lNI$.

\section{Conclusion and discussion}\label{sec:disc}

In this paper we studied the nearly-isotonic regression with respect to a
general partial order, with a weighted sum of squares, where some of the
weights can be equal to zero, and with an additional quadratic penalty. In
our opinion, the main result of the paper is Theorem \ref{thm:main}, where we
proved that this problem can be written as a generalized lasso problem with
a design matrix of full column rank and the same penalty matrix. Therefore,
the estimator can be computed by the algorithms developed for the generalized
lasso. Further, we derived the dual problem and wrote it as a sparse quadratic
program, which can be solved by standard solvers for sparse problems. Also,
we showed that for a sufficiently large penalty parameter the estimator
coincides with the regularized isotonic regression.

One of the possible directions is to study the degrees of freedom of the
estimator. In Theorem 2 of \cite{TT}, for a design matrix of full column rank
and Gaussian data with a covariance matrix proportional to the identity
matrix, an unbiased estimate of the degrees of freedom of the fitted values of
the generalized lasso is derived. In our Theorem \ref{thm:main} the design
matrix has full column rank, but the fitted values $X\hvbeta$ are not the
estimate of interest, and the data vector $\tvy$ depends on $\lNI$. Moreover,
even for Gaussian observations with a covariance matrix proportional to the
identity matrix, the covariance matrix of $\tvy$ is not of this form.
Therefore, the result of \cite{TT} does not apply directly, and it is
interesting to derive the degrees of freedom of the estimator $\hvbeta$
itself, in the spirit of Theorem 6 in \cite{Pastukhov}.

Another direction is to develop a path algorithm with respect to $\lNI$.
Since the data vector $\tvy$ in Theorem \ref{thm:main} depends on $\lNI$,
the path algorithm of \cite{TT} does not give it directly. In particular, it
is interesting to find the smallest value of $\lambda_{0}$ in
Section \ref{sec:exact}, for which the estimator coincides with the
regularized isotonic regression.

\end{document}